\documentclass[journal=jacsat,manuscript=article]{achemso}

\usepackage{graphicx}
\usepackage{caption}
\usepackage{subcaption}
\usepackage{amsmath}
\usepackage{amsbsy}
\usepackage{amsthm}
\usepackage{amsfonts}
\usepackage{float}
\usepackage{sidecap}
\usepackage{mathtools}
\usepackage{braket}
\usepackage{adjustbox}

\usepackage[utf8]{inputenc}
\usepackage[T1]{fontenc}
\usepackage{graphicx}
\usepackage{longtable}
\usepackage{wrapfig}
\usepackage{rotating}
\usepackage[normalem]{ulem}
\usepackage{amsmath}
\usepackage{amssymb}
\usepackage{float}
\usepackage{colortbl}
\usepackage{capt-of}
\usepackage{hyperref}
\usepackage{tabularx}

\begin{tocentry}

\includegraphics{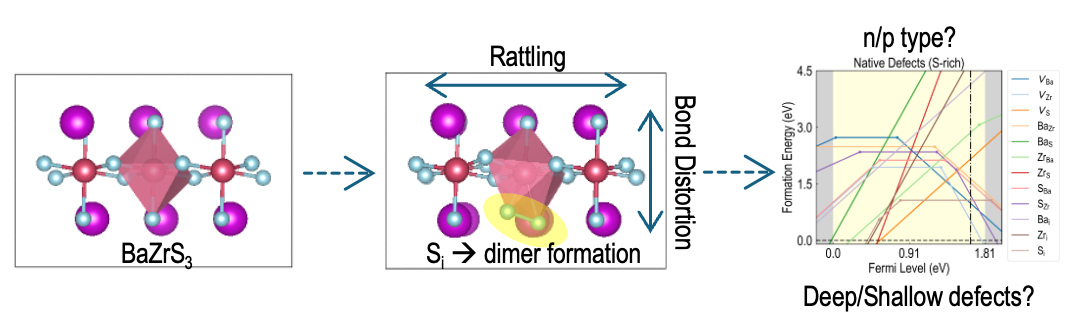}

\end{tocentry}

\author{Rushik Desai}
\email{desai224@purdue.edu}
\affiliation[]
{School of Materials Engineering, Purdue University, West Lafayette, IN 47907, USA}

\author{Shubhanshu Agarwal}
\email{agarw228@purdue.edu}
\affiliation[]
{School of Chemical Engineering, Purdue University, West Lafayette, IN 47907, USA}

\author{Kiruba Catherine Vincent}
\email{vincen24@purdue.edu}
\affiliation[]
{School of Chemical Engineering, Purdue University, West Lafayette, IN 47907, USA}

\author{Alejandro Strachan}
\email{strachan@purdue.edu}
\affiliation[]
{School of Materials Engineering, Purdue University, West Lafayette, IN 47907, USA}

\author{Rakesh Agrawal}
\email{agrawalr@purdue.edu}
\affiliation[]
{School of Chemical Engineering, Purdue University, West Lafayette, IN 47907, USA}

\author{Arun Mannodi-Kanakkithodi}
\email{amannodi@purdue.edu}
\affiliation[]
{School of Materials Engineering, Purdue University, West Lafayette, IN 47907, USA}

\title{Exploring the defect landscape and dopability of chalcogenide perovskite BaZrS$_3$}

\begin{document}

\begin{abstract}
BaZrS$_3$ is a chalcogenide perovskite that has shown promise as a photovoltaic absorber, but its performance is limited because of defects and impurities which have a direct influence on carrier concentrations. Functional dopants that show lower donor-type or acceptor-type formation energies than naturally occurring defects can help tune the optoelectronic properties of BaZrS$_3$. In this work, we applied first principles computations to comprehensively investigate the defect landscape of BaZrS$_3$, including all intrinsic defects and a set of selected impurities and dopants. BaZrS$_3$ intrinsically exhibits n-type equilibrium conductivity under both S-poor and S-rich conditions, which remains largely unchanged in the presence of O and H impurities. La and Nb dopants created stable donor-type defects which make BaZrS$_3$ even more n-type, whereas As and P dopants formed amphoteric defects with relatively high formation energies. This work highlights the difficulty of creating p-type BaZrS$_3$ owing to the low formation energies of donor defects, both intrinsic and extrinsic. Defect formation energies were also used to compute expected defect concentrations and make comparisons with experimentally reported values. Our dataset of defects in BaZrS$_3$ paves the path for training machine learning models to subsequently perform larger-scale prediction and screening of defects and dopants across many chalcogenide perovskites, including cation-site or anion-site alloys.
\end{abstract}

\section*{Introduction}
Increasing energy demands across the globe have positioned photovoltaics as one of the primary research focus areas among energy materials. Discovering novel absorber compositions that can yield better photovoltaic (PV) efficiencies while bringing down costs remains a challenge, and exploring material classes beyond traditional semiconductors has been a considerable thrust. Perovskites have gained attention as viable replacements for solar absorbers owing to their promising electronic and optoelectronic properties \cite{perovskite-1, perovskite-2}. Hybrid organic-inorganic perovskites (HOIPs) with the general formula ABX$_3$ (where X is a halogen anion) have been widely explored for high-efficiency solar absorption, but suffer from a number of challenges. There are well-documented stability issues that manifest as protonation in the presence of moisture and photo- or heat-induced phase segregation into halide-rich regions \cite{oihp, degradation-1, degradation-2}. HOIPs are largely composed of Pb at the B-site and thus face issues of toxicity and environmental safety. Replacement or partial substitution with ions such as Sn or Ge either lead to lower PV efficiencies or easy oxidation because of their variable oxidation states \cite{stability-perovskite-1, stability-perovskite-2, stability-perovskite-3}. Inorganic chalcogenide perovskites (ICPs) are an emerging sub-class currently being investigated for their exciting and tunable optoelectronic properties \cite{chalcogenide-1, chalcogenide-2, chalcogenide-3}. ICP compositions are typically non-toxic and exhibit superior aqueous and thermal stability compared to their halide counterparts\cite{thermal-stability, thermal-stability-1, phase-stability}. This family of materials also demonstrates considerable chemical versatility, positioning them as important candidates for next-generation renewable energy applications\cite{htdft-chalco}. \\

Structurally, ICPs conform to the canonical ABX$_3$ architecture, where S or Se occupies the X site, a divalent or trivalent cation is positioned at the A site, and a transition metal cation in a suitable oxidation state resides at the B site. Considerable attention has been directed towards the compound BaZrS$_3$ because of its high thermal (up to $600^{\circ}$ C) and aqueous stability \cite{bazrs3-1, bazrs3-2}. Due to its suitable electronic structure and optical absorption behavior, BaZrS$_3$ has found application in photovoltaics \cite{bazrs3-photovoltaics-1, bazrs3-photovoltaics-2, bazrs3-photovoltaics-sim}, photodetectors \cite{photodetector-1, bazrs3-1}, and photocatalysis \cite{catalysis-1, catalysis-2, catalysis-3}. Multiple synthesis routes have been proposed for crystalline BaZrS$_3$, including sulfurization of oxide precursors, solution processing, sputtering, and liquid flux-assisted crystal growth \cite{bazrs3-1, synthesis-1, synthesis-2, synthesis-3, synthesis-4, synthesis-5}. Various computational studies employing density functional theory (DFT) on bulk and surface structures of BaZrS$_3$ have verified its promising properties, positioning it as an important non-toxic energy harvesting material \cite{dft-1, dft-2, dft-3, dft-3, dft-4, dft-5, dft-6}. \\ 

Despite several numerical studies showing high photovoltaic (PV) efficiencies $>$ 38\%, only a single device has been made to date using BaZrS$_3$ which has shown poor efficiency \cite{bazrs3-scaps, bazrs3-photovoltaics-sim, device-1}. One of the possible causes for the poor performance is the material's inherent equilibrium electrical conductivity which is determined by the lowest energy donor- and acceptor-type native point defects and unintentional impurities in the system. The concentrations of these defects along with the electron and hole concentrations dictate whether the equilibrium Fermi level is around the middle of the band gap (intrinsic), closer to the valence band edge (p-type), or closer to the conduction band edge (n-type). Furthermore, deep defect states arising from native defects or impurities can increase the non-radiative recombination of electrons and holes and create absorption peaks at undesired energies. DFT computations provide a suitable way to systematically investigate defect properties and have been performed extensively to understand the stability of point defects in semiconductors and potentially identify shallow and deep-level states \cite{dft-defect, defects}. Multiple studies have also been published where DFT was used to study defect-related properties of BaZrS$_3$. \\

Meng et al. \cite{defect-meng} found that a S interstitial defect (S$_i$) in BaZrS$_3$ creates a deep mid-gap state. They also reported that Ti alloying at the Zr site could be a viable avenue to increase the PV efficiency of the material. Still, the alloyed composition is unstable due to its positive decomposition energy. Another theoretical study by Wu et al. \cite{defect-wu} discussed the defect landscape of BaZrS$_3$ where they also found S$_i$ to be a stable deep-level defect. Due to the formation of S-S dimers and S-S-S trimers, sulfur-based defects are stabilized. Another recent study by Yuan et.al.\cite{defect-yuan} delved deeper into the effect of S$_i$ and found that it creates a strong recombination center for charge carriers, which may further explain the lower device efficiencies. The defect studies performed indicate that the native defects have a strong influence on the electrical conductivity of BaZrS$_3$. A systematic analysis of the affect of chemical and synthesis conditions and extrinsic doping on the properties of interest is still missing. \\

Alloying studies have been performed to understand how composition engineering may help improve PV efficiencies. Sharma et al. used machine learning to identify suitable alloying to decrease the band gap of BaZrS$_3$ \cite{ml-dopants}. They trained crystal graph convolutional neural network (CGCNN) models \cite{cgcnn} on DFT data from a variety of alloy structures, including 38 different elements in specific concentrations, and subsequently used the models to make predictions on a wide range of alloying concentrations. They found that Ca at the Ba site and Ti at the Zr site can reduce the band gap and enhance PV efficiency. Sharma et al. also reported the synthesis of thin films of the optimized compositions. Ti alloying was also shown by Meng et al.\cite{defect-meng} to lead to deep defects. Ca substitution at the A site was considered superior, and it would cause a reduction of the band gap to $\sim$ 1.26 eV. Other alloying studies have also discussed Ti-Zr and Se-S mixing to tune the electrical properties \cite{bazrs3-photovoltaics-1, alloying-ti, alloying-se-1, alloying-se-2}. Most of the compositions are found to be unstable or to not drastically improve the electrical properties. BaZrS$_3$ still shows the best combination of stability and optoelectronic properties. Thus, understanding how functional dopants behave in this material relative to native defects would further help tune the properties. In this work, we comprehensively explore the defect landscape in BaZrS$_3$, determining formation energies and related properties of native and extrinsic defects using non-local hybrid DFT. \\

\begin{figure}[ht]
    \centering
    \includegraphics[width=0.95\textwidth]{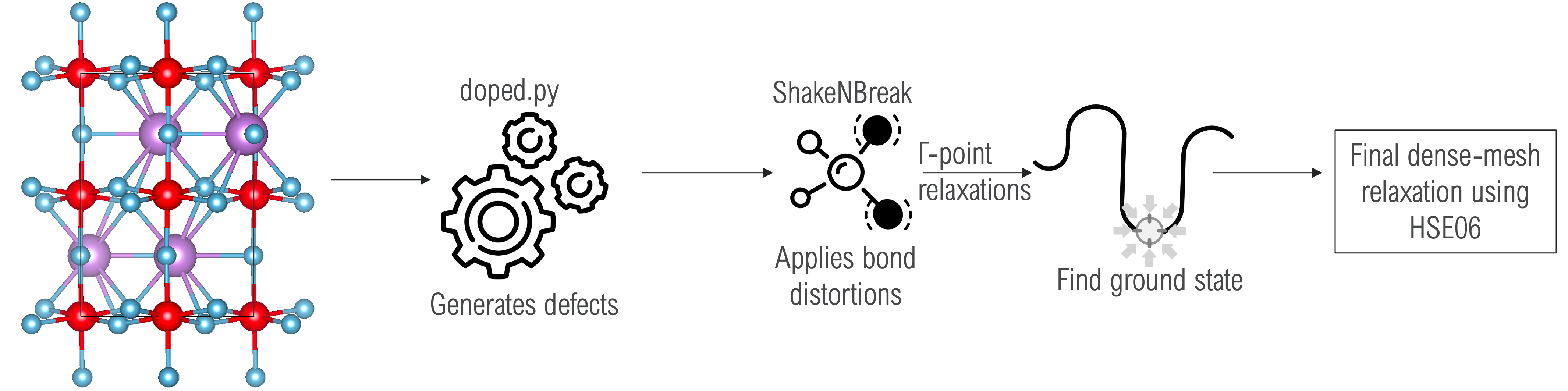} 
    \caption{DFT workflow for studying intrinsic and extrinsic point defects in BaZrS$_3$.}
    \label{fig:workflow}
\end{figure}

Modeling defect properties using DFT is a well-established approach \cite{defect-wu, defect-meng, ge-reviewer}; however, generating accurate defect structures can be challenging. While optimizing these structures, algorithms may become confined to local minima, failing to capture the correct ground state defect environment. In this study, we employed the ShakeNBreak \cite{shakenbreak} method as implemented in the ``Doped'' \cite{doped} package to generate defect structures by introducing external distortions into the system, which facilitates the exploration of the entire potential energy surface. A general overview of the workflow is presented in \textbf{Fig. \ref{fig:workflow}}. Using this approach, we computed the formation energies of a series of native point defects, impurities, and dopants in BaZrS$_3$ as a function of defect charge, Fermi level, and chemical potential conditions. We further used these energies to calculate values of equilibrium Fermi level and defect concentration. Our results help identify the lowest energy donor and acceptor defects and their expected concentrations, and understand how extrinsic defects would affect the electrical properties. This analysis subsequently helps make judgments on what type of doping may be necessary to make the equilibrium conductivity of BaZrS$_3$ more p-type. \\

\section*{Computational Details}

\begin{figure}[h]
\begin{center}
 \includegraphics[width = 0.9\textwidth]{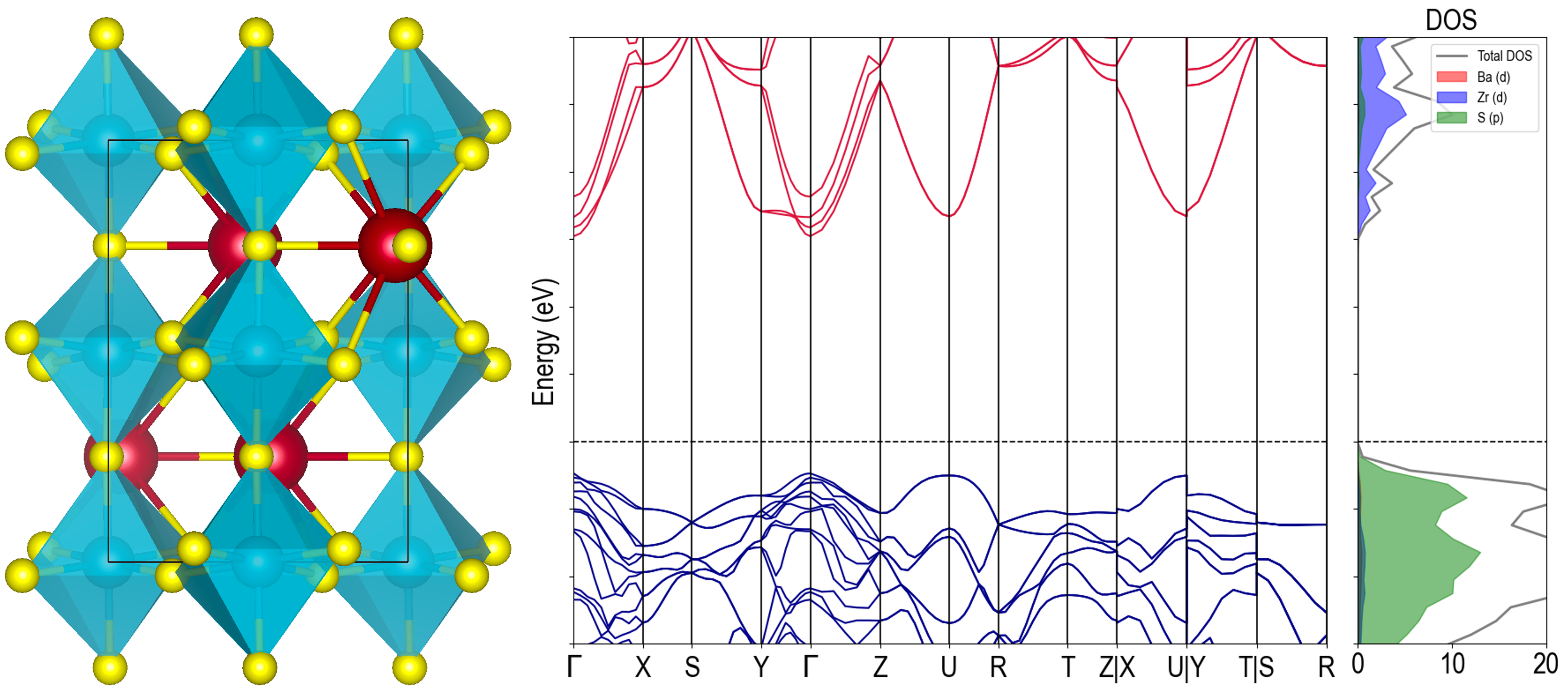}
  \caption{Optimized crystal structure, electronic band structure, and density of states of BaZrS$_3$ computed using HSE06. The red atoms are Ba, the center of blue polyhedra is Zr, and the yellow atoms are S.} \label{fig:band_structure}
  \end{center}
\end{figure}

All DFT computations were performed using the Vienna Ab initio Simulation Package (VASP)~\cite{vasp-1, vasp-2, vasp-3} with the Heyd-Scuseria-Ernzerhof (HSE06)~\cite{hse} non-local hybrid functional to accurately obtain exchange-correlation energies. The projector augmented-wave (PAW) method~\cite{paw} was employed to describe the electron-ion interactions. The initial orthogonal crystal structure, obtained from the Materials Project \cite{MP}, was first relaxed using the PBEsol functional ~\cite{pbesol} within the generalized gradient approximation (GGA) and subsequently relaxed using HSE06~\cite{hse}. The PBEsol functional is better parameterized for solids than other GGA functionals and enables quicker geometry optimization of BaZrS$_3$-related structures before the more expensive HSE06 functional is used to obtain electronic and defect properties more accurately, given the well-known errors in electronic levels from GGA alone.\cite{hse-validation-1, hse-validation-2, hse-validation-3} We computed the electronic band structure and found the band gap to be 1.76 eV, which closely matches what has been reported in the literature. \cite{defect-meng, defect-wu, defect-yuan, dft-3, dft-4}. The crystal structure, computed band structure, and density of states for BaZrS$_3$ are presented in \textbf{Fig. \ref{fig:band_structure}}. \\

Defects were introduced in a 2$\times$2$\times$1 supercell of BaZrS$_3$ containing 80 atoms by removing, inserting, or substituting atoms as necessary to create vacancies, interstitials, and substitutional defects. The Doped package \cite{doped} creates dozens of rattled and distorted defect configurations, which are initially fully optimized using only $\Gamma$-point calculations. Subsequently, denser mesh calculations using a 2$\times$2$\times$2 k-point mesh were performed to obtain the lowest energy structures and their energies. The general workflow is presented in \textbf{Fig. \ref{fig:workflow}}. Geometry optimization of all defect structures is performed in several charge states by explicitly removing or adding electrons to the system. \\

The total energies of the defect supercells in different charge states obtained from HSE06 are used to calculate the defect formation energies as follows:

\begin{equation}
    E_f = E_{\text{defect}} - E_{\text{bulk}} + \sum n_i \mu_i + q(E_{\text{VBM}} + E_{\text{F}}) + E_{\text{corr}}
    \label{dfe}
\end{equation}

Here, \(E_{\text{defect}}\) is the total energy of the supercell containing the defect in a charge state \textit{q}, \(E_{\text{bulk}}\) is the total energy of the pristine supercell, \(n_i\) represents the number of atoms of type \textit{i} added or removed to create the defect, \(\mu_i\) represents the chemical potential values of species \(i\), \(E_{\text{F}}\) is the Fermi level which ranges from the valence band maximum (VBM) to the conduction band minimum (CBM) computed for BaZrS$_3$, and \(E_{\text{corr}}\) is the charge correction energy applied to take into account the defect-image interactions \cite{fnv-1, fnv-2}. The correction energy is added to compensate for the self-interaction of a defect with its periodic image. This is done by considering the electrostatic potential energies of different atomic sites and examining the difference from the bulk structure to the defect structure, assuming the defect site is a point charge. This helps reduce the spurious interaction due to neighboring defect images and thus corrects the total energy. In this work, we apply the approach from Kumagai and Oba to determine the charge correction energy for any charged defect \cite{fnv-1, fnv-2}. The defect formation energy plots ultimately obtained from solving \textbf{Eqn. \ref{dfe}} help understand the relative likelihood of forming defects under different growth conditions, the p-type, intrinsic, or n-type nature of the equilibrium Fermi level, the expected defect concentrations as a function of temperature, and the ease or difficulty of doping the semiconductor one way or another. \\

All the chemical potential values required for \textbf{Eqn. \ref{dfe}} were obtained by considering the energies of the elemental standard states of each species (Ba, Zr, S, and the extrinsic dopants or impurities), the BaZrS$_3$ bulk phase, as well as all possible competing phases that may form. For instance, we considered that intrinsically, phases such as BaS, ZrS$_3$, and Ba$_3$Zr$_2$S$_7$ may form and thus restrict the range of possible energies where BaZrS$_3$ is in thermodynamic equilibrium. Similarly, phases such as Nb$_2$S$_3$ and ZrO$_2$ may form when extrinsic defects are in play. Structures of all elemental and competing phases were obtained from the Materials Project \cite{MP}, following which we further optimized them using the computational parameters described earlier to obtain their HSE06 energies at a consistent level of theory. A series of energy equations were solved to ensure that BaZrS$_3$ is in thermodynamic equilibrium and no competing phases may form; the resulting chemical potential energy diagram is pictured in \textbf{Fig. S1}. All chemical potential values are referenced to the energies of their elemental standard states. Only a very narrow set of Ba, Zr, and S chemical potential values satisfy all conditions. We use two extreme chemical potential conditions for calculating the defect formation energies, namely S-rich (or Zr-poor) and S-poor (or Zr-rich), respectively, referring to conditions where the chemical potential of S is close to the energy of S gas and the chemical potential of Zr is close to the energy of the ground state hexagonal phase of Zr. At either extreme, chemical potential values of all species are calculated by fixing the chemical potentials of S or Zr using their elemental energies.

Since synthesis studies \cite{synthesis-3} show that BaZrS${_3}$ is only synthesized at temperatures of $600^{\circ}$ C and higher, it would be useful to also determine defect concentrations at these temperatures. The defect concentrations and the equilibrium Fermi level are calculated in a self-consistent manner by applying charge neutrality conditions upon the computed concentrations of electrons, holes, donor-type defects, and acceptor-type defects. Electron (\textit{n$_0$}) and hole (\textit{p$_0$}) carrier concentrations are first estimated by using the electronic density of states based on the following equations: 

\begin{equation}
    n_0(E_f) = \int_{E_{CBM}}^{\infty} \frac{g(E)}{e^{(E-E_F)/k_BT} + 1} \, dE
\end{equation}

\begin{equation}
    p_0(E_f) = \int_{-\infty}^{E_{VBM}} \left(1 - \frac{1}{e^{(E-E_F)/k_BT} + 1}\right) g(E) \, dE
\end{equation}

Here, T is the temperature of interest, g(E) represents the density of states in the conduction band and valence band respectively for electrons and holes, k${_B}$ is the Boltzmann constant, E$_{CBM}$ is the conduction band minimum, and E$_{VBM}$ is the valence band maximum. For seamless calculation of the defect concentrations, we employed the ``py-sc-fermi'' method from J. Buckeridge et al. \cite{buckeridge, py-sc-fermi}, which is already encoded in the Doped package. Typically, defect concentrations are calculated using Arrhenius-type relationships involving the defect formation energy calculated for the most likely/relevant charge state and Fermi level. This treatment does not consider that most semiconductors are processed at high temperatures with much higher defect concentrations. Defects may get frozen in place and when quenched, the charge states get repopulated based on the annealed defect concentration. This has recently been observed in an experimental study where a large concentration of S vacancies was found at $600^{\circ} C$ \cite{new-defect}. After annealing, many vacancies still remained in the sample. To adequately account for this phenomenon, our methodology calculates concentrations at the annealing temperature using \textbf{Eqn. \ref{anneal_conc_our}}. Then, the value of $c_\text{d} (anneal)$ is substituted in \textbf{Eqn. \ref{defect_conc_our}} by freezing the concentration of defects at the annealing temperature. This leads to the correct relative concentration of \textbf{$c_\text{q} (operation)$} to be substituted in \textbf{Eqn. \ref{sc_fermi_our}}. The conventional methodology, on the other hand, involves solving only \textbf{Eqns. \ref{anneal_conc}, \ref{defect_conc_q}, and \ref{sc_fermi}}.

\begin{equation}
c_\text{d} = \exp\left(-\frac{g_{f,P}(T)}{k_B T}\right)
\label{anneal_conc}
\end{equation}

\begin{equation}
    c_{q} = c_d \frac{\exp\left(\frac{-g_{f,P}^q(E_F; T)}{k_B T}\right)}{\sum_q \exp\left(\frac{-g_{f,P}^q(E_F; T)}{k_B T}\right)}
    \label{defect_conc_q}
\end{equation}

\begin{equation}
    \rho(E_F) = \sum_{c_q,q} q[c_q] + p_0 - n_0 = 0
    \label{sc_fermi}
\end{equation}

The modified equations are presented thus:

\begin{equation}
c_\text{d} (anneal) = \exp\left(-\frac{g_{f,P}(T_{anneal})}{k_B T_{anneal{}}}\right)
\label{anneal_conc_our}
\end{equation}

\begin{equation}
    c_{q} (operation) = c_d (anneal) \frac{\exp\left(\frac{-g_{f,P}^q(E_F; T_{operation})}{k_B T_{operation}}\right)}{\sum_q \exp\left(\frac{-g_{f,P}^q(E_F; T_{operation})}{k_B T_{operation}}\right)}
    \label{defect_conc_our}
\end{equation}

\begin{equation}
    \rho(E_F) = \sum_{c_q (operation),q} q[c_q] + p_0 - n_0 = 0
    \label{sc_fermi_our}
\end{equation}

Here $T_{anneal}$ = 900 K ($\sim 600^\circ C$) is the annealing temperature, and $T_{operation}$ = 300 K is the operating temperature. q[c$_q$] represents the concentration of donor or acceptor defects, and g$_{f, P}$ is the defect formation energy obtained from DFT. Solving these equations simultaneously determines the system's equilibrium state, considering the inter-dependence of carrier concentrations, defect formation, and Fermi level position while satisfying charge neutrality. It is important to note that the total concentrations should be taken at the annealing temperature since, assuming the defects are not mobile, the concentrations of defects remain constant while quenching the material down to room temperature. \\

In the next few sections, we discuss the computed formation energies, temperature-dependent concentrations, and equilibrium Fermi levels for all native defects (vacancies \(\mathrm{V_{\text{Ba}}}\), \(\mathrm{V_{\text{Zr}}}\), and \(\mathrm{V_{\text{S}}}\), self-interstitials \(\mathrm{Ba_{\text{i}}}\), \(\mathrm{Zr_{\text{i}}}\), and \(\mathrm{S_{\text{i}}}\), and anti-site substitutions \(\mathrm{Ba_{\text{Zr}}}\), \(\mathrm{Ba_{\text{S}}}\), \(\mathrm{Zr_{\text{Ba}}}\), \(\mathrm{Zr_{\text{S}}}\), \(\mathrm{S_{\text{Ba}}}\), and \(\mathrm{S_{\text{Zr}}}\)), selected impurities O and H, and selected dopants La, Nb, As, and P, each at interstitial and substutional sites. \\

\section*{Results and Discussion}

\subsection*{Native Defects}

\begin{figure}[h]
\begin{center}
 \includegraphics[width = 0.9\textwidth, height= 0.5\textheight, keepaspectratio]{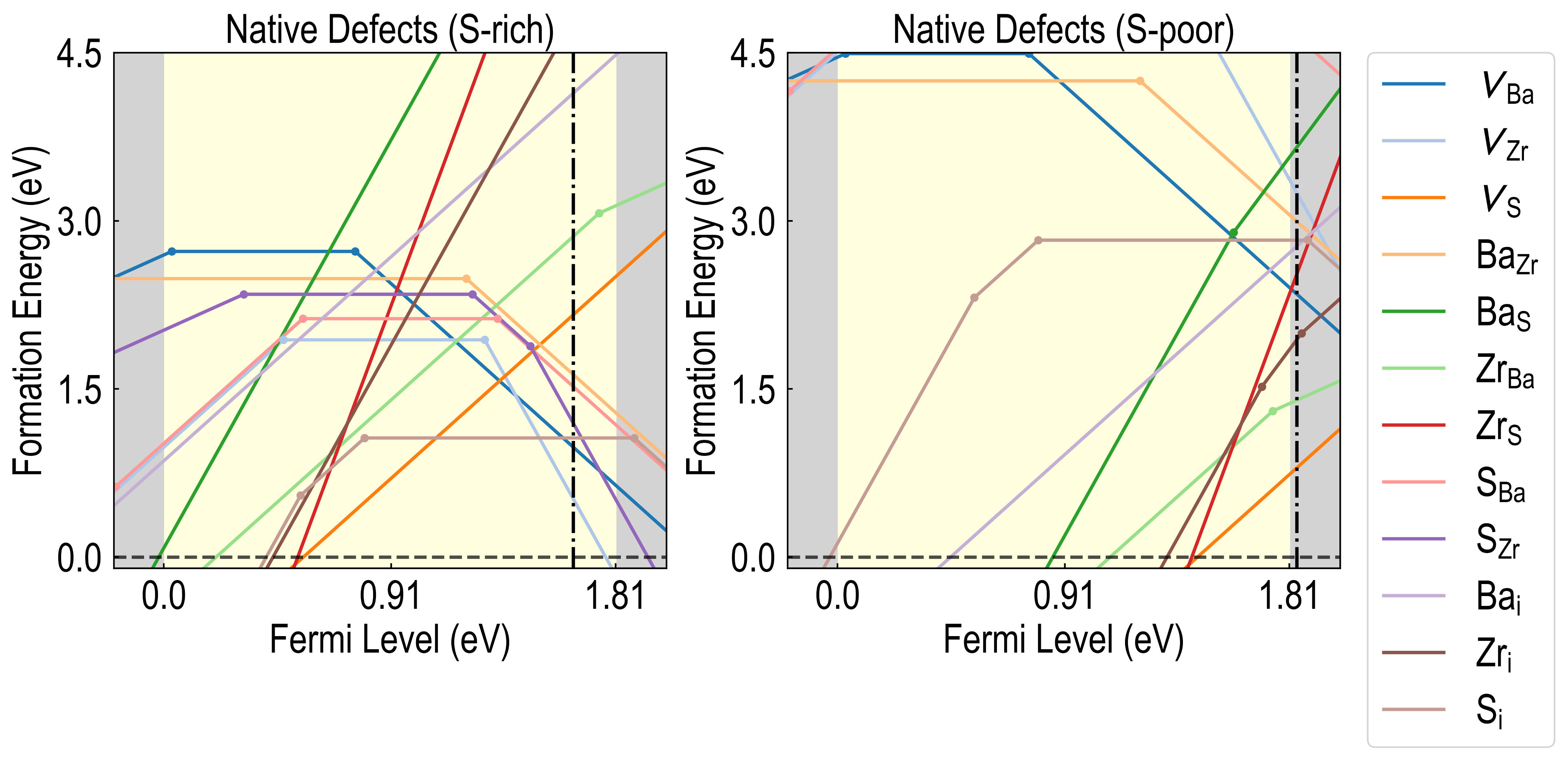}
  \caption{Defect formation energy plots for all native point defects in BaZrS$_3$, presented for S-rich and S-poor chemical potential conditions. The dotted-dash vertical line indicates the equilibrium Fermi level self-consistently calculated using an annealing temperature of 900 K and and operating temperature of 300 K.} \label{native_defects}
  \end{center}
\end{figure}

\textbf{Fig. \ref{native_defects}} shows defect formation energy plots for native point defects in BaZrS$_3$ under S-rich (or Zr-poor) and S-poor (or Zr-rich) chemical potential conditions. The diagrams are plotted as solutions of \textbf{Eqn. \ref{dfe}}, where the chemical potential values are obtained by solving thermodynamic equilibrium equations as described in the Computational Details, and the Fermi level is varied from the VBM to the CBM. The plot shows how the defect formation energy changes as a function of Fermi level, with a positive slope indicating a donor-type or n-type defect likely to enhance the electron concentration in the system, and a negative slope indicating an acceptor-type or p-type defect which would enhance hole concentration. For any given defect, a change in the slope indicates a charge transition level or a possible trap state, which, depending on its position in the band gap, can be termed a deep or shallow level. A comparison of the formation energies of different defects reveals which of them are most likely to exist and dominate over other defects under different chemical growth conditions or Fermi-level conditions (close to or far from either band edge). The computed defect formation energies also allow estimation of defect concentrations and thus a self-consistent calculation of the equilibrium Fermi level, as also explained in the Computational Details. \\

S-rich growth conditions have generally been utilized to synthesize BaZrS$_3$ \cite{synthesis-3} and would thus be more relevant towards understanding defect formation. Overall, our calculations reveal that BaZrS$_3$ contains several highly stable donor-type defects, which makes it a strongly n-type semiconductor, consistent with other DFT and experimental studies \cite{defect-meng, defect-wu, defect-yuan, new-defect}. Acceptors are relatively higher in energy than donors. \textbf{Fig. S2} shows plots of native defect formation energies with expanded y-axis ranges, picturing more defects. Under either chemical potential condition, certain native defects have low enough formation energies to exist in high concentrations, especially at Fermi levels closer to the VBM. Under S-poor conditions, donor defects formed by the vacancy V$_S$, interstitials Zr$_i$ and Ba$_i$, and anti-site defects Zr$_S$, Zr$_{Ba}$, and Ba$_S$, all show relatively low formation energies. Each donor defect shows shallow ionization levels close to or inside the CB. Cation vacancies and sulfur interstitial (S$_i$) show relatively higher energies, though S$_i$ creates deep mid-gap transition levels. This is consistent with previous theoretical studies where S$_i$ was shown to be a strong recombination center \cite{defect-yuan} that potentially annihilates generated carriers. \\

Under S-rich conditions, most of the donor defects have slightly higher energies, whereas acceptor defects such as V$_{Ba}$, Ba$_{Zr}$, S$_{Ba}$, and S$_{Zr}$ show lower energy compared to S-poor conditions. Interestingly, S$_i$ has a much lower formation energy here, and its deep mid-gap states become more consequential. The equilibrium Fermi level, as determined by the lowest energy acceptor and donor defects, is moderately n-type under S-rich conditions as opposed to very n-type under S-poor conditions. We also find deep transition levels shown by cation vacancies V$_{Ba}$ and V$_{Zr}$. It should be noted that even in S-rich conditions, the sulfur vacancy shows very low formation energy, which might indicate a Frenkel-type defect forming in the material. Furthermore, we find S-S dimer and S-S-S trimer formation in the optimized defect configurations for S$_{i}$. This type of S-cluster formation which helps passivate dangling bonds is likely the reason for the low energy of this defect. Wu et al. \cite{defect-wu} also made a similar observation. Atomistic representations of the dimer and trimer formation observed in our optimized S$_{i}$ defect configurations are presented in \textbf{Fig. S3}. \\

\begin{figure}[h]
\begin{center}
 \includegraphics[width = 0.8\textwidth, height= 0.5\textheight, keepaspectratio]{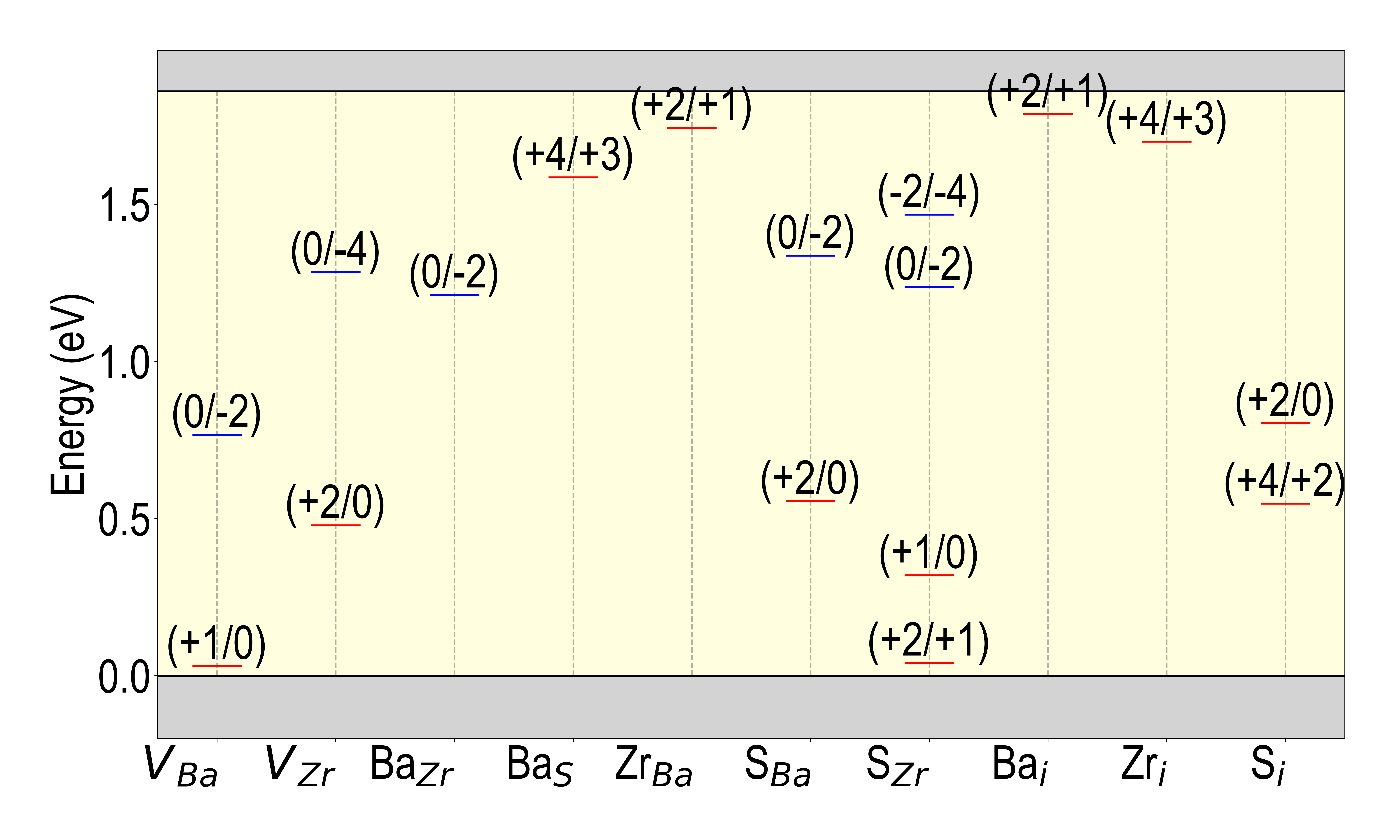}
  \caption{Computed charge transition levels for native defects in BaZrS$_3$. Red bars represent donor-type levels, and blue bars represent acceptor-type levels.} \label{CTLs}
  \end{center}
\end{figure}

\textbf{Fig. \ref{CTLs}} presents all the relevant charge transition levels (CTLs) corresponding to native defects in BaZrS$_3$. Many defects create mid-gap CTLs but are generally high in formation energy and thus not problematic, except for S$_i$. Deeper levels created by S$_i$ can lead to non-radiative recombination centers, which can exhaust electron-hole pairs by acting as trap states and thus impair the photovoltaic application. Vacancies V$_{Ba}$ and V$_{Zr}$ create deep 0/-2 and +2/0 levels in the band gap, whereas V$_S$ only creates a shallow donor level. Due to the low energy of V$_{S}$ in both growth conditions, the Fermi level is pushed to a more n-type region, which makes electrons the primary charge carriers. Anti-site substitutional defects Ba$_{Zr}$, S$_{Ba}$, and S$_{Zr}$ also show deep levels, while other anti-sites generally create shallow levels. While cation vacancies create shallow donor levels, S$_i$ creates two mid-gap CTLs corresponding to +4/+2 and +2/0 transitions. \\

Using \textbf{Eqn. \ref{anneal_conc_our}}, \textbf{\ref{defect_conc_our}}, and \textbf{\ref{sc_fermi_our}} along with the computed defect formation energies, we determined the self-consistent Fermi level to be 1.84 eV (from the VBM) for S-poor conditions. This value is larger than the band gap of 1.76 eV and thus inside the CB, which suggests that there would be an excess of free electrons as charge carriers, making the material readily conduct under S-poor conditions. Under S-rich conditions, the equilibrium  Fermi level is calculated to be 1.64 eV, indicating moderate to strong n-type conductivity. Overall, our native defect calculations agree with previous works that report similar transition levels and formation energies from DFT \cite{defect-yuan}. \\

\begin{figure}[h]
\begin{center}
 \includegraphics[width = \textwidth, height= 0.5\textheight, keepaspectratio]{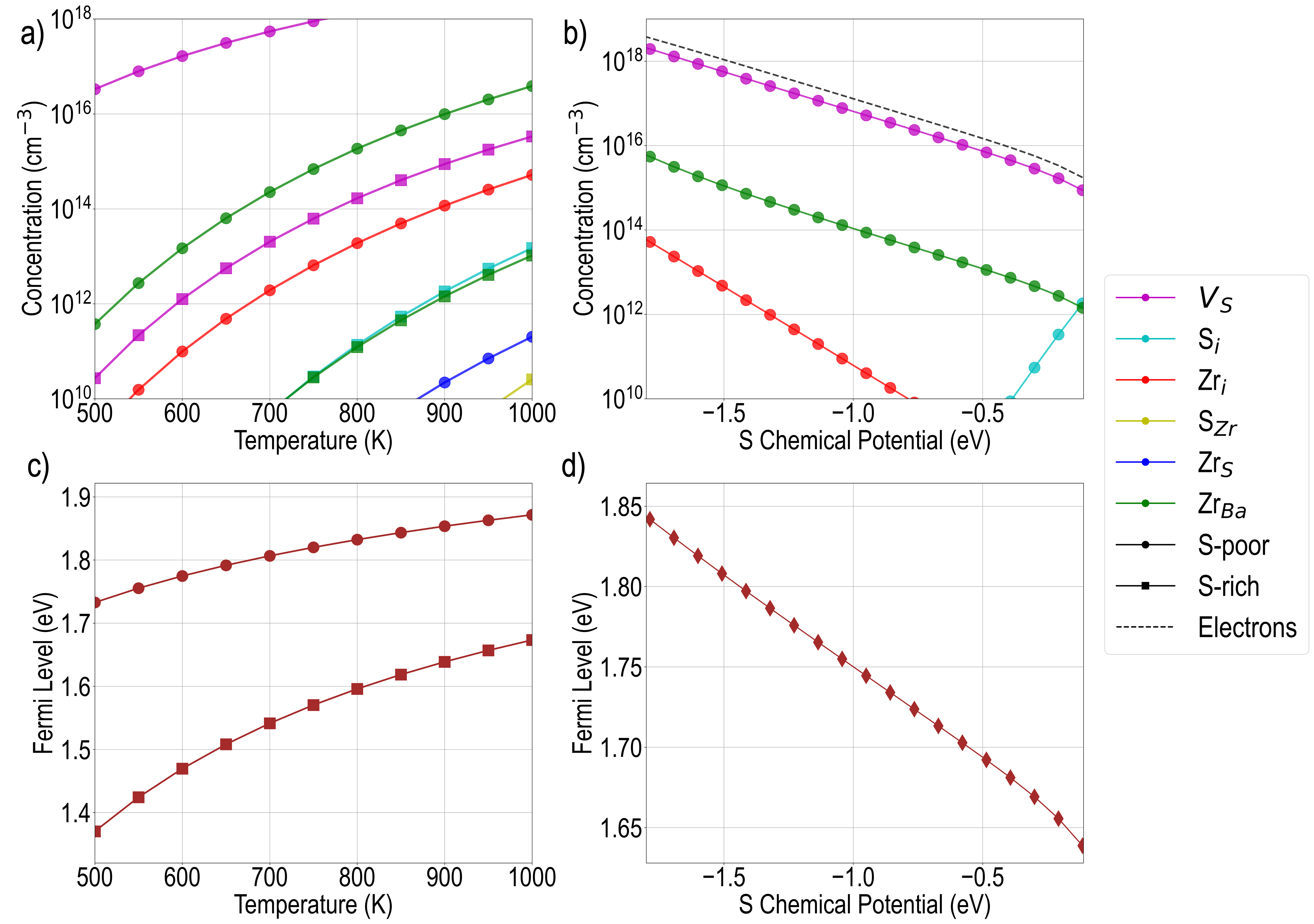}
  \caption{Defect concentrations, carrier concentrations, and equilibrium Fermi level plotted as functions of temperature and chemical potential. a) Concentrations plotted against annealing temperature. b) Concentrations plotted against chemical potential of S. c) Computed equilibrium Fermi level plotted against the annealing temperature for both S-poor and S-rich conditions. d) Equilibrium Fermi level plotted against the chemical potential of S.} \label{fig:fermi_conc}
  \end{center}
\end{figure}

To explore the effect of temperature on defect formation and related properties, we tuned the annealing temperature in \textbf{Eqn. \ref{anneal_conc_our}} from 500 K to 1000 K and examined how the defect concentrations change, keeping the operation temperature in \textbf{Eqn. \ref{defect_conc_our}} as the room temperature (300 K). \textbf{Fig. \ref{fig:fermi_conc}} shows plots for the defect concentration and self-consistent equilibrium Fermi level as functions of temperature and chemical potential conditions. Similar to what has been reported in previous studies 
\cite{defect-wu, defect-meng, defect-yuan}, V$_S$ shows the largest concentrations $\sim$ 10$^{19}$ cm$^{-3}$ as the chemical growth conditions become more S-poor. A recent experimental study by Aggrawal et al.\cite{new-defect} reported similar concentration ranges of $\sim$ 10$^{22}$ cm$^{-3}$ for sulfur vacancies, which validates our approach of freezing the defect concentrations. They also found that when the samples are annealed at 900 K, the dominant charge carriers are electrons with concentrations in the order of $\sim 10^{19} cm^{-3}$. Another study by Yang et al. found similar carrier-concentration ranges to ours at annealing temperatures of 600 K. \cite{carrier-conc-exp}\\

Due to the Arrhenius nature of the equation, we obtain very high defect concentrations at higher temperatures. Concentrations of Zr$_i$ and Zr$_{Ba}$ range between 10$^{14}$ and 10$^{16}$ cm$^{-3}$ and increase from S-rich to S-poor conditions. The equilibrium Fermi level rises from $\sim$ 1.5 eV to > 1.9 eV from S-rich to S-poor conditions. Since most low-energy defects are donor-type, the Fermi level gets pushed further inside the conduction band as the temperature increases. To understand how the chemical potential tunes the conductivity, we calculated the concentrations of defects and carriers (electrons and holes) at several intermediate chemical potential values and interpolated between the two extreme values. Since it is established that the material is predominantly n-type, the hole concentrations are quite low and are thus not pictured. We find that S-rich conditions are preferable for making the semiconductor less n-type or more intrinsic. In contrast, when conditions become more cation-rich/anion-poor, there is an increase in the concentration of both electrons and low-energy donor defects. It is evident that regardless of chemical growth conditions, the dominant donor defects will lead to n-type conductivity in BaZrS$_3$, which makes it important to study how impurities or dopants could help make it more p-type. \\

\subsection*{Impurities}

\begin{figure}[h]
\begin{center}
 \includegraphics[width = 0.9\textwidth, height= 0.5\textheight, keepaspectratio]{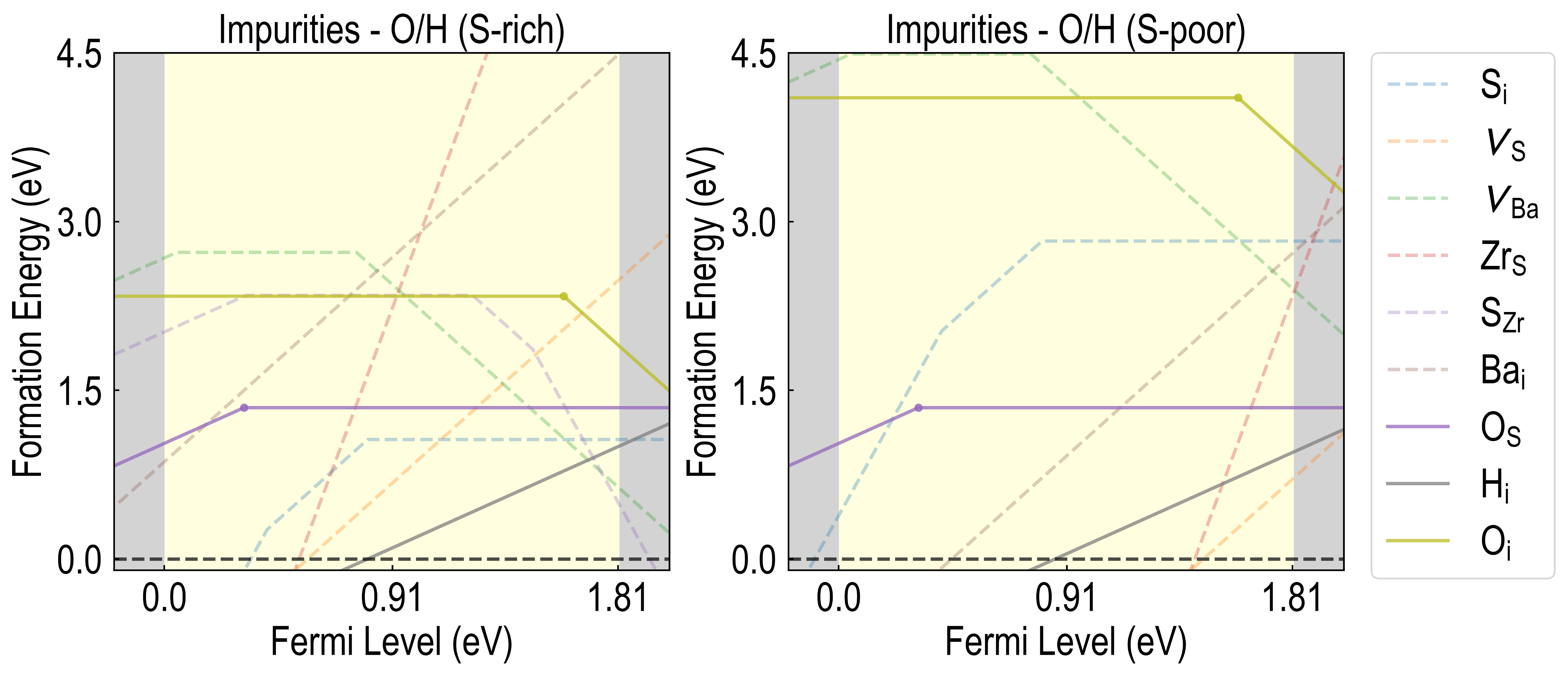}
  \caption{Computed formation energies of O and H impurities in BaZrS$_3$, compared with low energy native defects.} \label{O_H_defects}
  \end{center}
\end{figure}

Impurities introduced during synthesis could play a critical role in the functioning of a semiconductor. A recent study reported using ZrH$_2$ as a precursor for synthesizing BaZrS$_3$, leading to hydrogen impurities in the lattice \cite{synthesis-3}. Oxygen contamination on surfaces is also a common phenomenon affecting semiconductor manufacturing using thin film processing. Consequently, we computed the formation energies of O- and H-based point defects in BaZrS$_3$, including interstitials 
 O$_i$ and H$_i$, and substitutional defect O$_S$. \textbf{Figure \ref{O_H_defects}} shows their formation energies compared to the lowest energy native defects. O and H defects do not create deep levels and are reasonably stable under both S-rich and S-poor conditions. O$_S$ and O$_i$ both create electrically inactive stable neutral defects throughout the band gap region. H$_i$ creates a low energy donor-type defect and only transitions to acceptor-type (the H$^{1+}$/H$^{1-}$ transition level) within the CB. Overall, it is clear that while O and H defects can be expected in notable concentrations, they do not sufficiently dominate over the lowest energy native donors and will thus maintain the n-type conductivity in the compound. \\

\subsection*{Dopants}

\begin{figure}[h]
\begin{center}
 \includegraphics[width = 0.9\textwidth, height= 0.5\textheight, keepaspectratio]{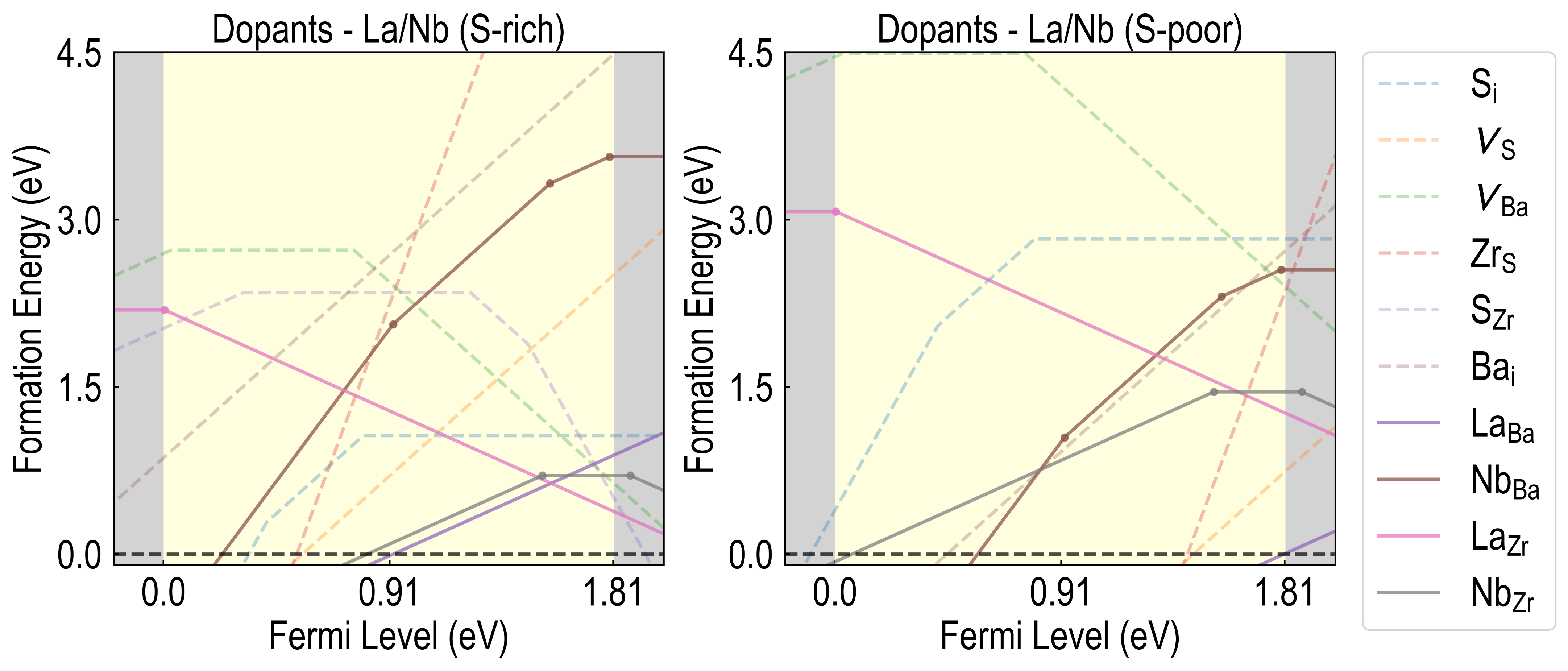}
  \caption{Computed formation energies of La and Nb dopants in BaZrS$_3$, compared with low energy native defects.} \label{fig:la-nb}
  \end{center}
\end{figure}

Here, we investigate a few selected cation-type and anion-type dopants in BaZrS$_3$ to understand if they dominate over native defects and thus help tune the electrical conductivity. La and Nb were chosen as substitutional dopants and studied at the Zr and Ba sites. Because of the similarity in sizes of La with Ba and Nb with Zr, these cations are hypothesized to be easy to incorporate as substitutional dopants, while their tendency to show multiple oxidation states may lead to either donor-type or acceptor-type defects. Size and oxidation state comparisons are provided in \textbf{Table \ref{dopants_comparison}}. \textbf{Fig. \ref{fig:la-nb}} shows the computed formation energies of La$_{Ba}$, La$_{Zr}$, Nb$_{Ba}$, and Nb$_{Zr}$ defects in comparison to the lowest energy native defects, for both S-poor and S-rich conditions. We find that La prefers to occupy the Ba site rather than the Zr site, with La$_{Ba}$ forming a very low energy donor-type defect as opposed to the higher energy acceptor La$_{Zr}$. La$_{Ba}$ slightly increases in energy from S-poor to S-rich conditions, but on this evidence, it will make the compound (along with low-energy native acceptors) just as n-type as before. Neither La$_{Ba}$ nor La$_{Zr}$ create any deep transition levels. \\

\begin{table}[]
\begin{tabular}{|l|l|l|l|l|}
\hline
Dopant & Host & Size (Host) & Size (Dopant) & Stable Charge (Dopant) \\ \hline
La     & Ba   & 215 pm      & 207 pm        & +3, +2                 \\ \hline
Nb     & Zr   & 175 pm      & 164 pm        & +5, +4, +3                     \\ \hline
As     & S    & 105 pm      & 119 pm        & -3                     \\ \hline
P      & S    & 105 pm      & 107 pm        & -3                     \\ \hline
\end{tabular}
\caption{Comparison between the covalent radii and known oxidation states of selected dopants and host atoms.}
\label{dopants_comparison}
\end{table}

Interestingly, the Nb dopant also prefers the Ba site instead of Zr under S-poor conditions, with both Nb$_{Ba}$ and Nb$_{Zr}$ being donor defects with relatively low energies closer to the VBM. Nb$_{Zr}$ becomes more stable than Nb$_{Ba}$ throughout the band gap region under S-rich conditions. While Nb$_{Zr}$ creates a shallow donor level, Nb$_{Ba}$ creates a deep mid-gap +3/+2 CTL, showing the tendency of Nb to exist in both 5+ and 4+ oxidation states. These deep levels could be problematic under S-poor conditions but not so much for S-rich conditions owing to higher energies. Similar to La, Nb will also keep the equilibrium conductivity very n-type. \textbf{Table S1} shows the self-consistent equilibrium Fermi levels calculated in the presence of individual dopants and compared with the values from native defects only. \\

\begin{figure}[h]
\begin{center}
 \includegraphics[width = 0.9\textwidth, height= 0.5\textheight, keepaspectratio]{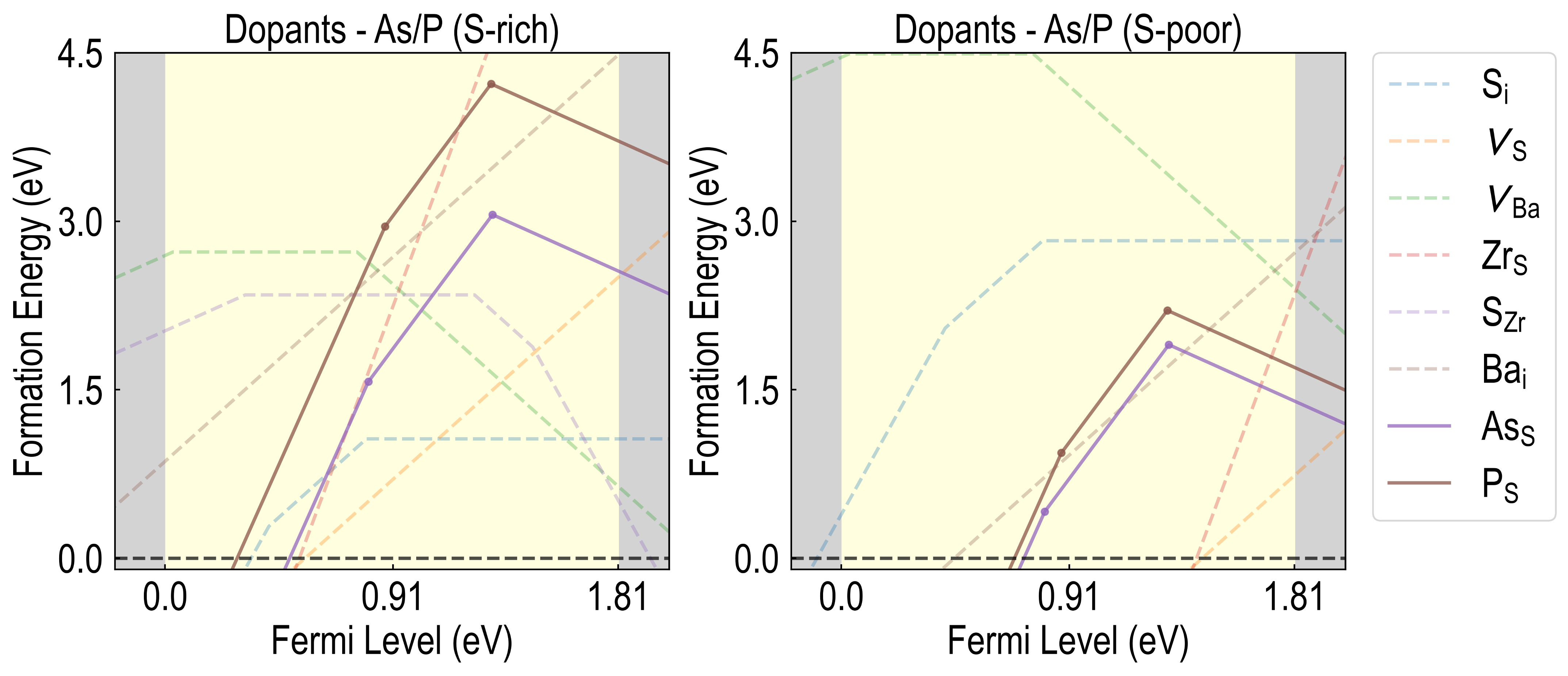}
  \caption{Computed formation energies of As and P dopants in BaZrS$_3$, compared with low energy native defects.} \label{fig:as_p}
  \end{center}
\end{figure}

Finally, we simulated As and P dopants as substitutional defects at the S site. Their computed formation energies are presented in \textbf{Fig. \ref{fig:as_p}} along with low energy native defects. We find that both As$_S$ and P$_S$ create donor-type defects for most of the band gap region and acceptor-type defects for a minority of it, and they both create deep mid-gap levels arising from +2/+1 and +1/-1 transitions. Their formation energies are generally higher than the native defects, meaning they are unlikely to change the equilibrium Fermi level, as shown in \textbf{Table S1}. Thus, doping with La, Nb, As, or P will only lead to n-type conductivity in BaZrS$_3$ due to either creating very low energy donor defects or higher energy defects that do not affect native defects. \textbf{Fig. S4} shows how the equilibrium Fermi level changes with temperature in the presence of different dopants, while \textbf{Table S2} shows the relevant charge transition levels computed for each dopant. \\

\clearpage

\section*{Perspective and Future Work}

Our calculations have shown that the extremely low energy (and high concentrations) of donor-type native defects (such as sulfur vacancies and interstitials) in BaZrS$_3$ makes its conductivity moderately n-type under S-rich conditions and very n-type under S-poor conditions. We further saw that certain dopants selected for their potential acceptor effect also lead to n-type conductivity. A much wider space of possible dopants must be systematically studied in BaZrS$_3$ to understand their effect on the electronic properties. Using the self-consistent Fermi level from \textbf{Eqn. \ref{sc_fermi}}, one could supply a dummy dopant atom in a particular charge state and calculate what specific concentration would be needed to cause a change in the material's electrical conductivity. Here, we performed this exercise using code implemented in the Doped package \cite{doped}. A dopant atom \textit{M} with charge \textit{r} would change the charge neutrality equation as follows:

\begin{equation}
    \rho(E_F, r[M_r]) = \sum_{c_q,q} q[c_q] + p_0 - n_0 + r[M_r]
    \label{dummy_dopant}
\end{equation}

\begin{figure}[h]
\begin{center}
 \includegraphics[width = 0.6\textwidth, height=0.6\textheight, keepaspectratio]{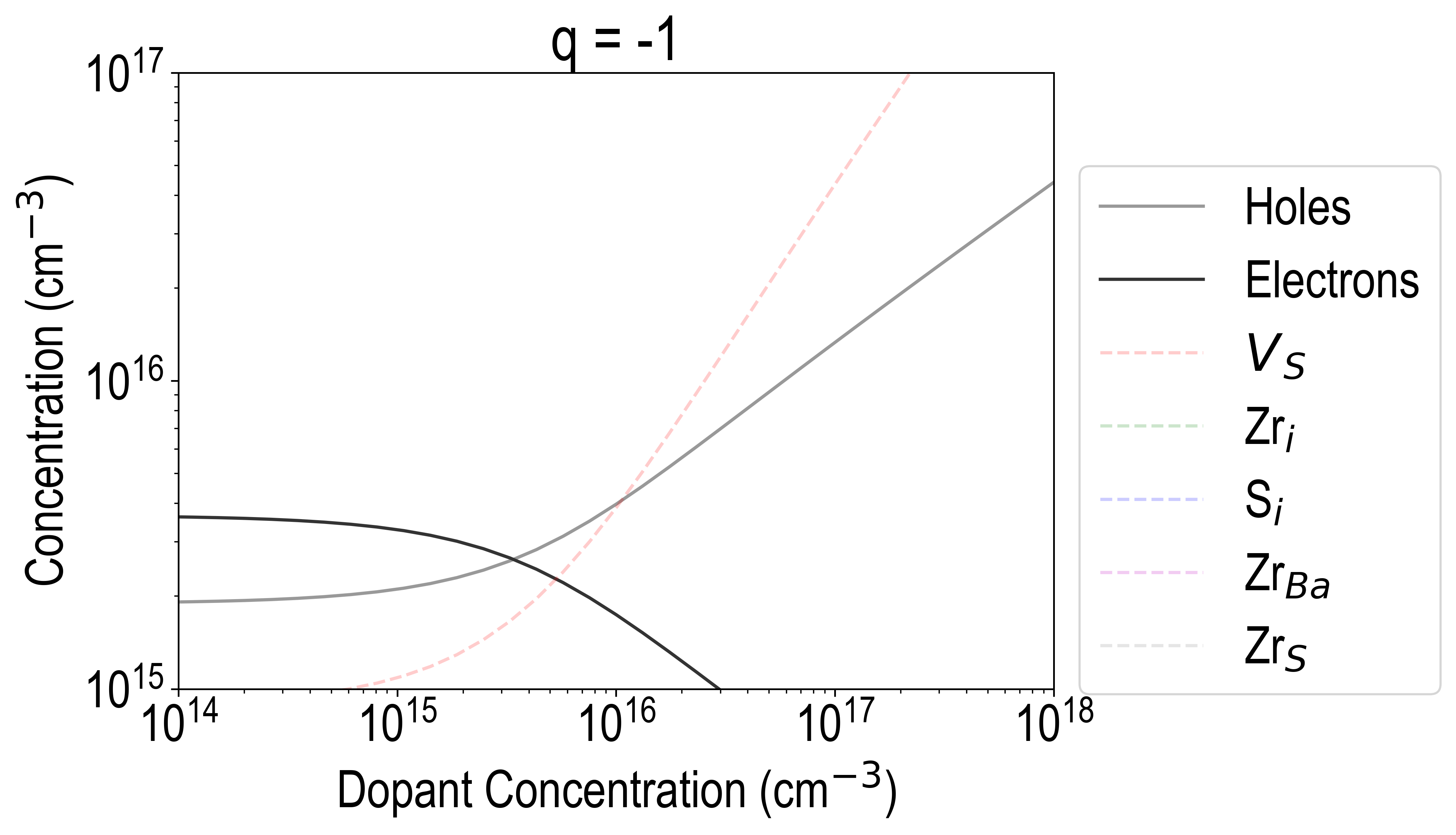}
  \caption{For a hypothetical defect in a 1- charge state, calculations show that p-type conductivity can be achieved in BaZrS$_3$ if the concentration of the dopant is > 10$^{16}$ cm$^{-3}$.} \label{fig:dummy_dopant_conc}
  \end{center}
\end{figure}

Knowing that we would need an acceptor-type dopant, we calculated the concentration required by the dopant in a -1 charge state. The above equation should yield the dopant concentration required to compensate for the effect of native defects on the material's electrical conductivity.  From \textbf{Fig. \ref{fig:dummy_dopant_conc}}, it can be seen that such a defect would be needed in a concentration of > 10$^{16}$ cm$^{-3}$ at the annealing temperature of 900 K to shift the Fermi Level to a p-type region by dominating over the native defects. \\

This motivates future work on high-throughput DFT computations for possible p-type dopants in BaZrS$_3$ selected from across the periodic table. It would also be important to look at defect complexes created by combinations of different dopants or with native defects to examine their relative energies and tendency to dominate over problematic single native defects. Furthermore, strategies must be explored to overcome the deep mid-gap states created by low-energy sulfur interstitial defects and reduce the possible non-radiative recombination of charge carriers. Given the expense of exploring dozens of possible rattled/distorted defect configurations in different charge states using hybrid functionals, machine learning approaches can help accelerate this screening \cite{defect-irena}. For instance, graph-based neural network models can help quickly predict the energies of defect structures and obtain ground state configurations. \cite{defect-irena, defect-habibur} \\

This study also highlights the importance of considering distorted structures when computing defect properties for realistic estimations of electrical conductivity, as evidenced by the match between computed defect concentrations and measured values \cite{new-defect}. BaZrS$_3$, while not prone to deep-level defects, shows low energy donor defects which lead to very high electron concentrations. This behavior aligns with observations in other chalcogenide compounds such as Cu$_2$SiSe$_3$ and Sb$_2$S$_3$ \cite{chalc-semicon-1, chalc-semicon-2}. The computational framework employed in this investigation will be extended to analyze other promising photo absorber materials, including BaHfS$_3$ and BaHfSe$_3$. A comparable high-throughput methodology was recently applied by us in a separate study examining zincblende-derived ternary and quaternary chalcogenide semiconductors, revealing the presence of low-energy anti-site and vacancy defects \cite{chalc-semicon-3}. \\

\section*{Conclusion}

We performed systematic DFT computations to study the properties of native and extrinsic point defects in the chalcogenide perovskite BaZrS$_3$. Based on the computed defect formation energies, we determined that certain native defects would form spontaneously under S-rich and S-poor conditions. Sulfur vacancies and interstitials both create low energy donor defects, with S$_{i}$ forming deep mid-gap trap states, which could potentially hamper the use of this perovskite as a solar absorber. Based on the defect concentrations and self-consistent Fermi levels computed from the formation energies, we find that BaZrS$_3$ is an n-type material owing especially to the low energy of the donor defects. O and H impurities show reasonably low formation energies but are electrically inactive compared to native defects. We also computed the formation energies of La and Nb dopants as substituents at Ba and Zr sites and As and P dopants as substituents at S site, which revealed that they predominantly form donor-type defects that are low in energy and would retain the n-type conductivity of BaZrS$_3$. Finally, we proposed a design objective to potentially change the material's conductivity to p-type by doping an element that can form an acceptor-type defect with a concentration greater than 10$^{16}$ cm$^{-3}$.   \\

\section*{Conflicts of Interest}
There are no conflicts to declare.

\section*{Data Availability}
All the DFT data generated in this work, scripts used, and .cif files of optimized structures are provided as part of the Supporting Information.
\begin{itemize}
    \item cifs.zip includes the CIF files for defect structures with different distortions.
    \item scripts.zip contains all scripts for plotting the data, along with JSON files storing relevant information about the defect analysis for different systems and chemical potential conditions.
    \item The Excel file (final\_dfe.xlsx) contains correction energies, chemical potentials, and DFT total energies for different defect systems.
\end{itemize}

\section*{Acknowledgements}
A.M.K. acknowledges support from the Materials Engineering Department at Purdue University. This research used resources from the Rosen Center for Advanced Computing (RCAC) clusters at Purdue and the Center for Nanoscale Materials (CNM) at Argonne National Laboratory. Work performed at the CNM, a U.S. Department of Energy Office of Science User Facility, was supported by the U.S. DOE, Office of Basic Energy Sciences, under Contract No. DE-AC02-06CH11357. \\

\bibliography{mybibfile}

\end{document}